# Design Optimisation of Power-Efficient Submarine Line through Machine Learning

Maria Ionescu[1], Amirhossein Ghazisaeidi[1], Jérémie Renaudier[1], Pascal Pecci[2] and Olivier Courtois[2]

*[1]Nokia Bell Labs, Route de Villejust, 96120 Nozay, France*
*[2]Alcatel Submarine Networks, Route de Villejust, 96120 Nozay, France*
maria.ionescu@nokia-bell-labs.com

**Abstract:** An optimised subsea system design for energy-efficient SDM operation is demonstrated using machine learning. The removal of gain-flattening filters employed in submarine optical amplifiers can result in capacity gains at no additional overall repeater cost. © 2020 The Author(s)

## 1. Introduction

Subsea cables, the backbone of the today's global optical communications infrastructure, comprise of multiple fibre pairs each carrying tens of Tbit/s. A particularity of these systems is that further capacity increases are limited by electrical power feeding. Spatial division multiplexing (SDM) has been proposed to continue improving the throughput under existing electrical power constraints [1,2], but the essential requirement for power optimisation remains a challenge [3,4]. Conventional subsea links utilise gain-flattening filters (GFF) after each amplifier for spectral equalisation. A typical GFF applied over the C-band corrects ~1-2 dB of gain ripple of an erbium-doped fibre amplifier (EDFA). In transoceanic subsea links composed of over a hundred of such amplifiers, the absence of GFFs results in the rapid accumulation of high gain fluctuations and large power excursions across the wavelength-division multiplexed (WDM) channels. However, in an SDM context, the GFF removal at a fixed periodicity, besides simplifying the architecture, could prove beneficial to improve power efficiency, leveraging the excess gain through revising the strategy for channel power allocation within the optical bandwidth [5,6]. In this work, we propose an optimised subsea system design by reducing the number of GFFs and exploiting the gain fluctuations thus incurred by adjusting the power per channel with reinforcement learning to increase throughput.

## 2. System Design

### 1.1. EDFA Machine Learning Model

Machine learning has been used to characterise an EDFA under arbitrary input signal power conditions, similarly to [7]. We measured its gain over 89 50-GHz spaced channels across the C-band (1530 nm-1645 nm), after varying the input power randomly between 0 and 11.5 dBm (with a maximum ripple of 6 dB) and the current (100-800 mA), collecting 6516 data sets. A neuronal network of a 90-nodes input layer (channel input powers and driving current), a 100-nodes hidden layer and an 89-nodes output layer (gain per channel). The model predicts the EDFA gain with a root mean square error of 0.16 dB across the bandwidth. We then fixed the driving current to 500 mA, to operate the amplifier with an average gain of 10 dB and a total output power (TOP) of 16 to 18 dBm. For the system considered herein, the nonlinear threshold was ~20 dBm, therefore our operating conditions fall within the SDM assumptions.

### 1.2. Power Optimisation Model

The proposed system design shown in Fig 1(a), employs a GFF to compensate the cumulative gain variations of the $N+1$ EDFAs in series and identically replicates the transmitted signal powers per channel $\underline{P}=[P_{1530},…,P_{1565}]$. There is no fibre following the GFF, to allow as large an attenuation profile excursion as possible. This configuration pattern (i.e. N+1 EDFAs and a single GFF) is repeated $M$ times to achieve a minimum target distance of 13,400 km. In order to leverage on the cumulative gain fluctuations of the series of gain-unequalised EDFAs, reinforcement learning is employed to perform the optimisation of the signal power profile $\underline{P}$, targeting a maximisation of the system throughput. The model, represented in Fig 1(b), is implemented as a 3-layer neuronal network of 89, 128, 128 nodes respectively, and trained over 100 episodes. Each episode starts with a random spectrum as input onto which it iteratively takes up to 300 actions from a set of 128 possibilities. An action consists of pre-emphasising (or not) by 0.5 dB randomly selected ~5 nm sub-bands of the signal spectrum, while ensuring a constant 10 dB amplifiers gain and a TOP within the linear regime (16-18dBm). The network is trained using a policy-gradient approach, the Adam optimiser and a $3·10^{-4}$ learning rate and outputs rewards associated with each possible action as the log-probability of capacity improvement. Therefore, for a given configuration pattern (Fig 1(a): $M=1$), the network finds an optimised spectral shape, to maximise the capacity. Finally, we extend the transmission distance and apply the optimised signal to the full trans-Pacific system model. Unlike conventional systems, here the SNR is wavelength-dependent and varies with the unfiltered cumulative ASE. The analytical model we used to compute the SNR [8] pessimistically assumes constant nonlinear interference across the entire bandwidth, equal to that of the central channel. Given the operation in the strongly linear regime dictated by SDM, this could not a be limitation to the results and conclusions presented herein.

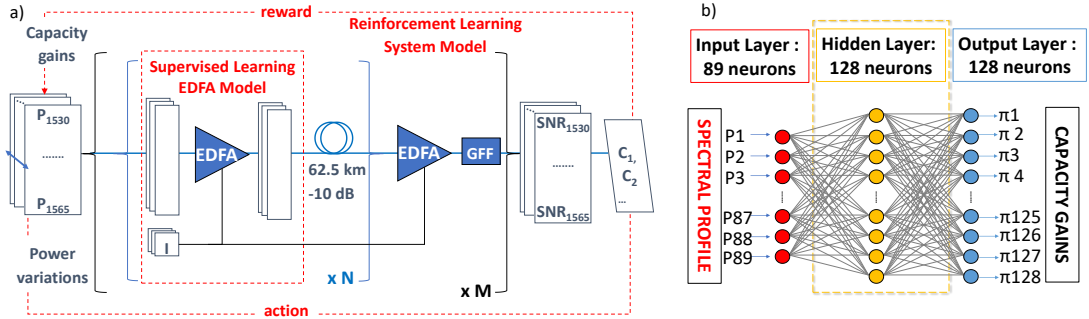

Fig. 1. (a) Model diagram of heterogeneous system design, comprising amplifiers both with and without GFF and (b) representation of neuronal network used for reinforcement learning.

### 3. Results

We analyse the capacity benefit resulting from a reduced GFF spacing and spectral optimisation. The benchmark is a conventional system design comprising of 250 fibre spans of 53.7 km (8.6 dB loss) and EDFAs operating at 10dB gain, each followed by an ideal GFF. The same EDFA ML model was used in both conventional and GFF-scarce design approaches. The reinforcement learning neural network finds optimum launch powers with excursion between 17-21 dBm, thus adjusting against the impact of cumulative ASE. When the spectrum is reshaped only every other amplifier, we can transmit 36.2 Tb/s compared to 31.3 Tb/s in a conventional system. However, the required number of amplifiers to attain 13,400 km increases significantly to 430 (72% higher). Despite the removal of GFFs enabling longer subsequent span lengths of 62.7 km at constant EDFA gain, we opted out of fibre at the output of the GFFs. Therefore, the number of amplifiers required is reduced as the filter spacing is increased. We found that a filter frequency of 7 required approximately the same number of repeaters (252), while still obtaining a capacity gain of 3.5 Tb/s compared to the conventional system.

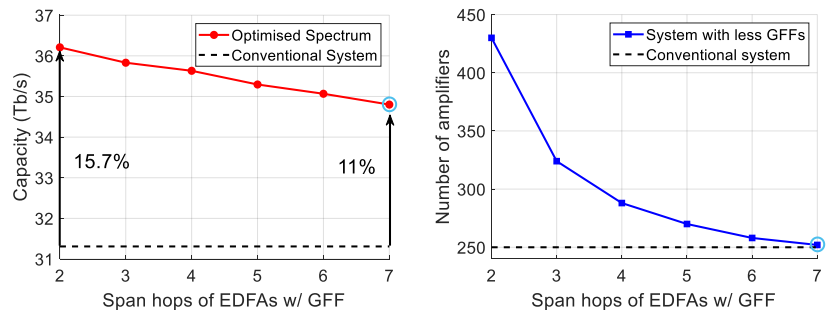

Fig. 2. Optimised capacity (left) and associated number of amplifiers required (right) given a GFF frequency.

### 4. Conclusions

We have demonstrated an alternative approach to subsea systems design, focusing on energy efficiency as an enabler of SDM operation, achieved by a reduction in number of GFFs, which perform spectral equalisation of multiple EDFAs. Machine learning techniques were utilized to both model a conventional EDFA and to find an optimal spectral shape of the transmitted C-band signal that would maximise the throughput. An 86% reduction in the number of GFFs can enable an 11% capacity increase, for no significant increase in number of amplifiers.